\documentclass[manuscript]{aastex}

 \shorttitle{Nuclear obscuration of H$_{2}$O maser galaxies} \shortauthors{Zhang et al.}

\begin{document}

\title{On the nuclear obscuration of H$_{2}$O maser galaxies}

\author{J. S. Zhang \altaffilmark{}\affil{Center for astrophysics, GuangZhou university, GuangZhou, 510006, China} \email{jszhang@gzhu.edu.cn}}
\and
\author{C. Henkel \altaffilmark{}}\affil{Max-Planck-Institut f\"ur Radioastronomie, Auf dem H\"ugel 69, D-53121 Bonn, Germany}
\and
\author{Q. Guo, H. G. Wang, J. H. Fan\altaffilmark{}}
 \affil{Center for astrophysics, GuangZhou university, GuangZhou, 510006, China}

\begin{abstract}
 To shed light onto the circumnuclear environment of 22 GHz ($\lambda$ $\sim$ 1.3\,cm) H$_{2}$O maser galaxies, we have analyzed some of their
 multi-wavelength properties, including the far infrared luminosity (FIR), the luminosity of the $[\rm O\,{III}]\lambda5007$ emission line, the nuclear
 X-ray luminosity, and the equivalent width of the neutral iron $K\alpha$ emission line (EW (K$_{\alpha}$)). Our statistical analysis includes a total of
 85 sources, most of them harboring an active galactic nucleus (AGN). There are strong anti-correlations between EW ($K_{\alpha}$) and
 two ``optical thickness parameters'', i.e. the ratios of the X-ray luminosity versus the presumably more isotropically radiated
 $[\rm O\,{III}]$ and far infrared (FIR) luminosities. Based on these anti-correlations, a set of quantitative criteria, EW ($K_{\alpha}$)$>$300\,eV,
 $L_{2-10\,keV}$$<$2\,$L_{\rm [O\,{III}]}$ and $L_{\rm FIR}$$>$600\,$L_{2-10\,keV}$ can be established for Compton-thick nuclear regions. 18 H$_{2}$O
 maser galaxies belong to this category. There are no obvious correlations between the EW (K$_{\alpha}$), the $[\rm O\,{III}]$ luminosity and the
 isotropic H$_{2}$O maser luminosity. When comparing samples of Seyfert 2s with and without detected H$_{2}$O maser lines, there seem to exist
 differences in EW (K$_{\alpha}$) and the fraction of Compton-thick nuclei. This should be studied further. For AGN masers alone, there is
 no obvious correlation between FIR and H$_2$O maser luminosities. However, including masers associated with star forming regions, a linear correlation
 is revealed. Overall, the extragalactic FIR-H$_2$O data agree with the corresponding relation for Galactic maser sources, extrapolated by several
 orders of magnitude to higher luminosities.
\end{abstract}


\section{Introduction}           

 Thanks to a large number of dedicated surveys during the last 15 years, the number of galaxies known to host 22\,GHz ($\lambda$$\sim$1.3\,cm)
 H$_2$O masers has increased tenfold and 85 sources at distances larger than those of the Magellanic Clouds have been reported so far to
 exhibit H$_2$O maser emission (e.g., Braatz \& Gugliucci 2008, Darling et al. 2008, Greenhill et al. 2008). Some of them are AGN-related, while others
 are found in off-nuclear star forming regions. Among the masers clearly identified as AGN-related, a large fraction ($\sim$40\%) have been identified as
  ``disk-maser'' candidates (Kondratko et al. 2006). Their maser spots are associated with central, typically parsec sized molecular accretion disks and
  maser line spectra show ``high velocity features" (red-shifted and blue-shifted features), in addition to the systemic velocity components. The study
  of disk-maser sources has become a very important subject, permitting mass estimates of supermassive black holes, distance estimates of galaxies, and
  providing a perspective to improve the accuracy of the Hubble constant and to constrain the equation of state for the elusive dark energy
  (e.g. Braatz et al. 2009).

 Observations show that among AGN masers, H$_{2}$O maser spots locate preferentially in the nuclear regions of Seyfert 2 or LINER galaxies and
 most of them are heavily obscured ($N_{\rm H}$$>$10$^{23}$cm$^{-2}$; Braatz et al. 1997, Madejski et al. 2006, Zhang et al. 2006, Greenhill et al.
 2008). The nuclear X-ray source is generally believed to heat the gas to temperatures suitable for 22\,GHz H$_{2}$O maser emission
 (Neufeld et al. 1994), which is supported by a relation between maser luminosity, the unabsorbed intrinsic nuclear X-ray luminosity, and the mass of the
 black hole (Kondratko et al. 2006, Su et al. 2008).

  For the obscured nuclear regions H$_{2}$O masers and X-rays provide, unlike optical data, deeply penetrating views. The X-ray
  absorption along the line-of-sight to the nucleus can provide through spectral model fitting important information on the nature of
  the circumnuclear environment. With modern X-ray telescopes, high quality X-ray spectra have been obtained for more than 30 H$_{2}$O maser
 galaxies. Based on an analysis of such spectra, the nuclear column densities of maser host galaxies were investigated by Zhang et al. (2006) and
 Greenhill et al. (2008). While these studies have shown that nuclear H$_{2}$O masers are mostly found in environments with high column density,
 it is still open whether H$_{2}$O masers preferentially arise from Compton-thick ($N_{\rm H}$$>$10$^{24}$cm$^{-2}$) AGN.

   To reduce limitations and uncertainties when modeling X-ray spectra, here we try to provide additional constraints to evaluate line-of-sight
  column densities. The iron $K_{\alpha}$ emission line ($\sim$6.4\,keV) is the most prominent line in the X-ray spectra of AGN. It is believed to be
  produced either by transmission through the absorbing material (Leahy \& Creighton 1993) or via the process of X-ray scattering/reflection by the cold
  iron in the nuclear accretion disk (e.g., Lightman \& White 1988; Fabian et al. 2000) or torus (Ghisellini et al. 1994). So it provides useful
  information to constrain the column density that absorbs the continuum and, in some cases, to distinguish between Compton-thick and -thin sources.
  Flat hard X-ray spectra and high equivalent widths (EW (K$_{\alpha}$)) of the iron line were found in highly absorbed sources and are
  generally used to identify Compton-thick nuclei (e.g., Matt 1997). However, currently there are no quantitative criteria on EW (K$_{\alpha}$) yet as a
  probe of the gas absorption.

  Unlike the X-ray absorption, $[\rm O\,{III}]\lambda 5007$ and FIR emission have often been used as comparatively isotropic indicators of the
  intrinsic nuclear power (e.g, Mulchaey et al. 1994, Alonso-Herrero et al. 1997, Shu et al. 2007). The forbidden $[\rm O\,{III}]\lambda 5007$
  line emission originates in the narrow line region (NLR), which is assumed to be isotropic for both type 1 and type 2 systems.
  Due to possible shielding effects in the torus, Netzer et al. (2006) proposed that the $[\rm O\,{II}]\lambda 3727$ line is a
  better tracer than $[\rm O\,{III}]$ for its considerably larger emission region. However, this still needs to be checked, since the
  $[\rm O\,{II}]$ line can also arise from high mass star formation. And their results are based on a high-luminosity high-redshift AGN sample. For
   the less luminous, relatively nearby AGN as those of our maser sample, the $[\rm O\,{III}]\lambda 5007$ luminosity has been
  extensively used as an indicator of the intrinsic AGN power (e.g., Mulchaey et al. 1994, Alonso-Herrero et al. 1997, Maiolino \& Rieke 1995,
  Heckman et al. 2005, Risaliti et al. 1999, Bassani et al. 1999, Panessa et al. 2006, Lamastra et al. 2009). The FIR emission, produced on linear scales
  much larger than the nuclear torus, should be free of biases caused by the viewing angle and should be highly isotropic (e.g., Mulchaey et al. 1994).
  So even without a detailed knowledge of the X-ray spectrum, a comparison of the observed X-ray luminosity with those of the more isotropic tracers of
  nuclear power can help us to determine the gas absorption along the line of sight toward the AGN (e.g., Bassani et al. 1999).

 In the following, X-ray, $[\rm O\,{III}]$, and FIR data are collected for those H$_{2}$O maser galaxies which are located farther than the
 Magellanic Clouds and which are known to host 22\,GHz H$_{2}$O masers. The data are analyzed to determine gas column densities along the line of sight
 toward the AGN and to derive properties related to activity in highly obscured and therefore particularly elusive Compton-thick nuclear environments.

\section{Data}

  Data related to the galaxies with detected H$_{2}$O maser emission, located at larger distances than the Magellanic Clouds, are compiled in
  Table\,1. Among the 85 published H$_{2}$O maser sources, there are 66 AGN masers (63 ``megamasers'' with isotropic luminosities $L_{H_{2}O}>10\,
  L_{\odot}$ and 3 ``kilomasers'' with $L_{H_{2}O}<10\,L_{\odot}$). The megamasers are classified as such because of their luminosity. While most of them
 have not yet been studied in detail, all thoroughly investigated megamasers are not separated by more than a few pc from the line of sight to the AGN
 of their parent galaxy. Ten kilomasers are related to star formation regions and another nine kilomasers are still awaiting
 interferometric observations to investigate their nature. All activity types of these H$_{2}$O maser galaxies are listed in Table\,1 and their dominant
 types (following Bennert et al. 2009) are used for the statistics (for details, see Table\,2). Most of the AGN maser sources are Seyfert 2 galaxies or
 LINERs.

\begin{deluxetable}{llcccccccc}
\tabletypesize{\scriptsize}
\tablecaption{Physical Parameters of Extragalactic H$_{2}$O Maser
Sources$^{*}$.} \tablewidth{0pt} \tablehead{ \colhead{Source} &
\colhead{Type} & \colhead{Tel.} & \colhead{EW ($K_{\alpha}$)} &
\colhead{$F_{X}$} & \colhead{Ref.} & \colhead{F$_{FIR}$} &
\colhead{H$_{\alpha}$/H$_{\beta}$} & \colhead{$F_{[\rm OIII]}$} &
\colhead{Ref.} }

\startdata

\textbf{NGC\,17}   &Sy2,LIRG, H\,II &  &                       &                &       &1.33  &     &         &       \\
NGC\,23            &LINER,LIRG, H\,I&  &                       &                &       &1.09  &5.89 & 37.9    &Ho97   \\
\emph{IC\,10}      &                &  &                       &                &       &3.65  &12.5 & 442.85  &Ho97   \\
NGC\,235A          &Sy1             &  &                       &                &       &1.1   &     &         &       \\
NGC\,253           &Sy2,SBG,H\,II   &  &                       &                &       &73.56 &     &         &       \\
NGC\,262(Mrk\,348) &Sy2             &A &$212^{+68}_{-72}$      & 482            &Awa00  &0.34  &6.02 & 177     &Dah88  \\
                   &                &G &$230^{+120}_{-140}$    & 1270           &Bas99  &      &     &         &       \\
IRAS\,F01063-803   &                &  &                       &                &       &0.37  &     &         &       \\
NGC\,449(Mrk\,1)   &Sy2             &  &                       &                &       &0.32  &5.86 & 236.51  &Dah88  \\
NGC\,520           &SBG,H\,II       &  &                       &                &       &2.86  &     &         &       \\
\emph{NGC\,598}    &H\,II           &  &                       &                &       &67.71 &4.55 & 0.24    &Ho97   \\
\textbf{NGC\,591}  &Sy2             &X &$2200^{+700}_{-600}$   &20              &Gua05b &0.28  &     & 178     &Whi92  \\
NGC\,613           &Sy2,H\,II       &  &                       &                &       &2.6   &     &         &       \\
IC\,0184           &Sy2,H\,II       &  &                       &                &       &      &     &         &       \\
NGC\,1052          &Sy2,LINER       &B &$180^{+80}_{-90}$      & 400            &Ter02  &0.14  &2.82 & 13.3    &Dah88  \\
\textbf{NGC\,1068} &Sy2,Sy1         &X &$1200\pm500$           & 462            &Cap06  &25.01 &7.00 & 6780    &Dah88  \\
                   &                &A &$1210^{+260}_{-280}$   & 350            &Bas99  &      &     &         &       \\
NGC\,1106          &Sy2             &  &                       &                &       &0.19  &     &         &       \\
Mrk\,1066          &Sy2             &C &$1120^{+850}_{-650}$   & 23             &Shu07  &1.14  &8.51 & 514     &Whi92  \\
\textbf{NGC\,1320} &Sy2             &X &$2200^{+440}_{-430}$   &496.2           &Gre08  &0.29  &     &         &       \\
\textbf{NGC\,1386} &Sy2             &X &$1800^{+400}_{-300}$   &$27\pm5$        &Gua05b &0.72  &5.7  & 1020    &Sto89  \\
                   &                &A &$7600^{+8900}_{-5000}$ &  20            &Bas99  &      &     &         &       \\
IRAS03355+0104     &Sy2             &  &                       &                &       &0.13  &     &         &       \\
\emph{IC\,342}     &Sy2,H\,II       &  &                       &                &       &8.29  &7.69 & 3.4     &Ho97   \\
MG J0414+0534      &QSO1            &  &                       &                &       &      &     &         &       \\
\textbf{UGC\,3193} &                &  &                       &                &       &0.36  &     &         &       \\
UGC\,3255          &Sy2             &  &                       &                &       &0.19  &     &         &       \\
Mrk\,3             &Sy2             &X &$610^{+30}_{-50}$      & 590            &Bia05  &0.57  &6.67 & 4610    &Whi92  \\
                   &                &B &$650^{+182}_{-182}$    & 650            &Cap99  &      &     &         &       \\
                   &                &A &$997^{+300}_{-307}$    & 650            &Bas99  &      &     &         &       \\
\emph{NGC\,2146}   &H\,II           &  &                       &                &       &12.54 &11.1 & 30.47   &Ho97   \\
VII ZW\,73         &Sy2             &  &                       &                &       &0.21  &     &         &       \\
NGC\,2273          &Sy2             &X &$2200^{+400}_{-300}$   &$69^{+16}_{-12}$&Gua05b &0.7   &6.92 & 164     &Whi92  \\
                   &                &A &$1040^{+440}_{-460}$   &125             &Ter02  &      &     &         &       \\
\textbf{UGC\,3789} &                &  &                       &                &       &0.26  &     &         &       \\
Mrk\,78            &Sy2             &  &                       &                &       &0.15  &6.50 & 242.76  &Dah88  \\
J0804+3607         &QSO2            &  &                       &                &       &      &     &         &       \\
\emph{He\,2-10}    &SBG             &C &              &9.56$^{+0.63}_{-0.64}$&J\"{u}r05 &2.40  &     &         &       \\
\textbf{2MASX\,J08362280}&Sy2       &  &                       &                &       &0.17  &     &         &       \\
Mrk\,1210          & Sy2,Sy1        &C &$\sim$188              &840             &Zha09  &0.36  &5.20 & 580     &Ter91  \\
                   &                &X &130                    &970             &Gua02  &      &     &         &       \\
                   &                &B &$108^{+50}_{-65}$      &930             &Ohn04  &      &     &         &       \\
                   &                &A &$820^{+360}_{-430}$    &160             &Awa00  &      &     &         &       \\
\textbf{NGC\,2639} &Sy1.9           &A &$1490^{+11110}_{-1270}$& 25.3           &Ter02  &0.43  &4.16 & 4       &Ris99  \\
NGC\,2782          &Sy1,SBG         &C & 990                   &$\sim$30        &Zha06  &0.91  &6.67 & 62.19   &Ho97   \\
NGC\,2824(Mrk\,394)&Sy?             &  &                       &                &       &0.12  &     &         &       \\
SBS\,0927+49       &LINER           &  &                       &                &       &0.29  &     &         &       \\
\textbf{NGC\,2960} &LINER           &  &                       &                &       &0.24  &     &         &       \\
UGC\,5101          &Sy1.5,LINE      R,LIRG&X&$410^{+270}_{-240}$    & 8.1       &Ima03  &1.06  &     & 191     &Kim95  \\
NGC\,2979          &Sy2             &  &                       &                &       &0.2   &     &         &       \\
NGC\,2989          &H\,II           &  &                       &                &       &      &     &         &       \\
\emph{NGC3034}     &SBG,H\,II       &  &                       &                &       &96.49 &25.0 & 1615.17 &Ho97   \\
\textbf{NGC\,3079} &Sy2,LINER       &X &$1480\pm500$           & 33             &Cap06  &5.5   &25.0 & 92      &Ho97   \\
                   &                &B &$2400^{+2900}_{-1500}$ &37$\pm$8        &Iyo01  &      &     &         &       \\
\textbf{Mrk\,34}   &Sy2             &  &                       &                &       &0.43  &10.5 & 204.19  &Dah88  \\
\emph{NGC\,3359}   &H\,II           &  &                       &                &       &0.70  &     &         &       \\
\textbf{IC\,2560}  &Sy2             &X &2320$^{+180}_{-170}$&38.8$^{+1.8}_{-5.1}$&Til08 &4.29  &     &$>40$    &Ris99  \\
                   &                &C &2770$\pm$490       &38.4$^{+21.1}_{-4.6}$&Mad06 &      &     &         &       \\
\textbf{NGC\,3393} &Sy2             &X &$1400\pm800$           &$9^{+6}_{-4}$   &Gu05a  &0.33  &4.12 & 316     &Dia88  \\
                   &                &A &$3500\pm2000$          & 40             &Bas99  &      &     &         &       \\
NGC\,3556          &H\,II           &  &                       &                &       &7.29  &7.14 & 2.24    &Ho97   \\
ARP\,299(NGC\,3690)&                &X,C&$422^{+262}_{-288}$   & 43.7           &Bal04  &9.43  &5.88 & 35.6    &Ho97   \\
                   &                &B &$636^{+236}_{-270}$    &                &Del02  &      &     &         &       \\
NGC\,3735          &Sy2             &  &                       &                &       &1.02  &6.31 & 33      &Ho97   \\
\emph{Antennae}    &SBG             &C &                       &                &       &      &     &         &       \\
\textbf{NGC\,4051} &Sy1.5           &X &$240\pm40$             & 627            &Cap06  &1.32  &3.33 & 59.99   &Ho97   \\
NGC\,4151          &Sy1.5           &X &$300\pm30$             & 4510           &Cap06  &1.11  &3.45 &1695.39  &Ho97   \\
                   &                &A &101$\pm$5              &$\sim$20000     &Wea01  &      &     &         &       \\
\emph{NGC\,4214}   &SBG             &C &                       & 243            &Har04  &1.81  &     &         &       \\
\textbf{NGC\,4258} &Sy1.9,LINER     &X &$27\pm20$              & 837            &Cap06  &5.08  &9.12 & 262     &Hec80  \\
                   &                &A &250$\pm$100            & 300            &Bas99  &      &     &         &       \\
NGC\,4293          &LINER           &  &                       &                &       &0.58  &7.69 & 5.95    &Ho97   \\
\textbf{NGC\,4388} &Sy2             &X &$440\pm90$             & 762            &Cap06  &1.44  &5.50 &374$\pm$50&Bas99 \\
                   &                &A &$732^{+243}_{-191}$    & 1200           &Bas99  &      &     &          &      \\
NGC\,4527          &LINER,H\,II     &  &                       &                &       &3.80  &     &          &      \\
\textbf{ESO\,269-G012}&Sy2          &  &                       &                &       &0.23  &     &          &      \\
NGC\,4922          &Sy2,LINER       &  &                       &                &       &0.58  &7.14 & 33.03    &Kim95 \\
\textbf{NGC\,4945} &Sy2             &C & 1300                  & 500            &Don03  &41.38 &     &$>40$     &Ris99 \\
                   &                &B & $\sim$1300            & 540            &Gua00  &      &     &          &      \\
                   &                &A &$850\pm160$            & 350            &Bas99  &      &     &          &      \\
NGC\,5194          &Sy2,H\,II       &X &$986\pm210$            & 48             &Cap06  &6.62  &8.33 & 228      &Ho97  \\
                   &                &A &$910^{+350}_{-360}$    & 91.9           &Ter02  &      &     &          &      \\
\emph{NGC\,5253}   &SBG,H\,II       &C &                       & 29.9           &J\"{u}r05&3.16&     &          &      \\
Mrk\,266(NGC\,5256)&Sy2,LIRG,SBG    &B & 575                 & 56             &Ris00  &0.84  &5.92 & 44.33    &Dah88 \\
NGC\,5347          &Sy2             &C &$1300\pm500$           & 22             &Lev06  &0.27  &     & 114      &Tra01 \\
\textbf{NGC\,5495} &Sy2,H\,II?      &  &                       &                &       &0.27  &     &          &      \\
\textbf{Circinus}  &Sy2             &C &$2250^{+260}_{-300}$   & 1400           &Smi01  &26.11 &19.1 & 6970     &Bas99 \\
NGC\,5506(Mrk\,1376)&Sy1.9          &X &$86^{+24}_{-10}$       & 5800           &Bia03  &1.06  &7.20 & 333      &Lum01 \\
                   &                &A &$150\pm30$             & 8380           &Bas99  &      &     &          &      \\
NGC\,5643          &Sy2             &X & 500                   & 84             &Gua04  &2.59  &6.40 & 662      &Whi92 \\
                   &                &A &$1800^{+800}_{-960}$   & 130            &Bas99  &      &     &          &      \\
\textbf{NGC\,5728} &Sy2,H\,II       &C &$1100^{+320}_{-270}$   & 133            &Shu07  &0.96  &5.96 & 761      &Sto95 \\
                   &                &C &1130                   &                &Zha06  &      &     &          &      \\
\textbf{UGC09618NED02}&LINER,H\,II  &  &                       &                &       &0.92  &     &          &      \\
\textbf{NGC\,5793} &Sy2             &  &                       &                &       &0.6   &     &          &      \\
NGC\,6240          &Sy2,LINER       &C &$2400^{+800}_{-700}$   &170             &Pta03  &2.18  &17.2 &135$\pm$20&Kim95 \\
                   &                &A &$1580^{+380}_{-350}$   &190             &Bas99  &      &     &          &      \\
\textbf{NGC6264}   &Sy2             &  &                       &                &       &      &     &          &      \\
NGC\,6300          &Sy2             &X &$148\pm18$             &$2160\pm100$    &Mat04b &2.34  &     & 320      &Lum01 \\
\textbf{NGC\,6323} &Sy2             &  &                       &                &       &      &     &          &      \\
ESO\,103-G035      &Sy2,Sy1         &A &$173^{+50}_{-116}$     & 907            &Tur97  &0.37  &6.31 & 112      &Po196 \\
IRAS\,F19370-013   &Sy2,H\,II       &  &                       &                &       &0.26  &     &          &      \\
\textbf{3C\,403}   &FR$\rm II$      &C &244$\pm$20             &                &Kra05  &0.1   &     &          &      \\
\textbf{NGC\,6926} &Sy2,H\,II       &  &                       &                &       &0.65  &15.6 & 21.98    &Kim95 \\
AM\,2158-380NED02  &Sy2,RG          &  &                       &                &       &      &     &          &      \\
\textbf{TXS\,2226-184}&LINER        &  &                       &                &       &      &     &          &      \\
NGC\,7479          &Sy2,LINER       &  &                       &                &       &0.83  &10.0 & 37.9     &Kim95 \\
IC\,1481           &LINER           &  &                       &                &       &0.12  &     &          &      \\
\enddata

 \tablecomments{85 published extragalactic H$_{2}$O maser sources with available physical parameters are listed (10 masers arise in
 star forming regions marked by italics and 29 out of 66 AGN-masers are potential disk-masers, their source names are presented in boldface).
 Recently reported masers, not being part of the 78 sources listed by Bennert et al. (2009), are one type I quasar from Impellizzeri et al. (2008), two
 sources (NGC\,17 and NGC\,1320) from Greenhill et al. (2008), and four new masers related with star formation (He\,2-10, Antennae, NGC\,4214,
 NGC\,5253) from Darling et al. (2008).}

 \tablenotetext{Column\,1:}{Extragalactic H$_{2}$O maser host galaxies;}
 \tablenotetext{Column\,2:}{Type of nuclear activity. SBG: StarBurst Galaxy; Sy1, Sy1.5, Sy1.9, Sy2: Seyfert types; LINER: Low-Ionization Nuclear
 Emission Line Region; LIRG: Luminous-Infrared Galaxy; FR\,II: Fanarov-Riley Type II radio galaxy; NLRG: Narrow-Line Radio Galaxy; RG: Radio Galaxy;
 H\,II: classified as a H\,II region; QSO1 and QSO2: type 1 and 2 Quasars. References: Zhang et al. (2006); Kondratko et al. (2006) and NED;}
 \tablenotetext{Column\,3:}{X-ray telescope--A: {\it ASCA; B: \it BeppoSAX}; C: {\it Chandra}; X: {\it XMM-Newton};}
 \tablenotetext{Columns\,4\&5:}{The EW ($K_{\alpha}$) of the Fe line (eV) and the 2-10\,keV observed X-ray flux (in units of
 10$^{-14}$ erg\,s$^{-1}$cm$^{-2}$);}
 \tablenotetext{Column\,6:}{References for Col. 4\&5 --- Awa00: Awaki et al. 2000; Bal04: Ballo et al. 2004; Bas99: Bassani et al. 1999;
   Bec04: Beckmann et al. 2004; Bia03: Bianchi et al. 2003; Bia05: Bianchi et al 2005; Cap99: Cappi et al. 1999; Cap06: Cappi et al. 2006;
   Del02: Della Ceca et al. 2002; Dia88: Diaz et al. 1988; Don03: Done et al. 2003; Gua00: Guainazzi et al. 2000a; Gua02: Guainazzi et al. 2002;
   Gua04: Guainazzi et al. 2004; Gua05a: Guainazzi et al. 2005a; Gua05b: Guainazzi et al. 2005b; Ima03: Imanishi et al. 2003;
   Iwa02: Iwasawa et al. 2002; Iyo01: Iyomoto et al. 2001; Jen04: Jenkins et al. 2004; Kra05: Kraft et al. 2005; Lev06: Levenson et al. 2006;
   Mad06: Madejski et al. 2006; Mat01: Matt et al. 2001; Mat04a: Matt et al. 2004a; Mat04b: Matsumoto et al. 2004b;
   Ohn04: Ohno et al. 2004; Pta03: Ptak et al. 2003; Ris00: Risaliti et al. 2000; Smi96: Smith \& Done 1996; Smi01: Smith \& Wilson 2001;
   Ter02: Terashima et al. 2002; Tur97: Turner et al. 1997; Wea01: Weaver et al. 2001; Zha06: Zhang et al. 2006; Zha09: Zhang et al. 2009;}
 \tablenotetext{Column\,7:}{FIR flux in units of 10$^{-9}$\,erg\,s$^{-1}$\,cm$^{-2}$;}
 \tablenotetext{Column\,8:}{The Balmer increment line intensity ratio H$_{\alpha}$/H$_{\beta}$;}
 \tablenotetext{Column\,9:}{Extinction corrected $[\rm O\,{III}]\lambda 5007$ line flux in units of $10^{-14}\,erg\,s^{-1}\,cm^{-2}$;}
 \tablenotetext{Column\,10:}{References for Col. 8\&9 --- Bas99: Bassani et al. 1999; Dah88: Dahari \& De Robertis 1988; Hec80: Heckman et al. 1980;
   Ho97: Ho et al. 1997; Kim95: Kim et al. 1995; Lum01: Lumsden et al. 2001; Pol96: Polletta et al. 1996; Ris99: Risaliti et al. 1999;
  Shu07: Shu et al. 2007; Sto89: Storchi-Bergmann \& Pastoriza 1989; Sto95: Storchi-Bergmann et al. 1995; Ter91: Terlevich et al. 1991;
  Tra01: Tran 2001; Whi92: Whittle 1992;}

\end{deluxetable}

\begin{table}
 \caption[]{Activity types of H$_2$O maser host galaxies (79 sources with available type)}
\begin{scriptsize}
\begin{center}
\begin{tabular}{l c c c c}
\hline

Type$^{a)}$  &SF-masers$^{b)}$ &Kilomasers of unknown origin$^{c)}$&AGN-masers$^{d)}$& the whole sample    \\

\hline
Seyfert 2    &     0           &        4                          &    45           &  49                 \\
LINER        &     0           &        2                          &     8           &  10                 \\
Inter Sy.    &     0           &        1                          &     2           &  3                  \\
Seyfert 1    &     0           &        0                          &     2           &  2                  \\
SBG          &     5           &        1                          &     0           &  6                  \\
H\,II        &     4           &        1                          &     1           &  6                  \\
FR\,II       &     0           &        0                          &     1           &  1                  \\
Quasar       &     0           &        0                          &     2           &  2                  \\

\hline

\end{tabular}
\end{center}
 a) Types from NED: Inter Sy., intermediate Seyfert types; LINER, low-ionization nuclear emission line region; SBG, StarBurst Galaxy;
 FR\,II, Fanaroff-Riley type II radio galaxy; H\,II, classified as H\,II region galaxy;
 b) SF-masers, H$_{2}$O masers associated with off-nuclear star formation regions;
 c) kilomasers without known type;
 d) AGN-masers, H$_{2}$O masers associated with AGN.
\end{scriptsize}
\end{table}

 Besides the type of nuclear activity, we collected multi-wavelength data and parameters for all of the 85 extragalactic
 H$_{2}$O maser sources, including the EW ($K_{\alpha}$) of the iron  emission line, the observed X-ray (2-10\,keV) flux F$_{X}$, the FIR flux
 F$_{FIR}$, the observed H$_{\alpha}/H_{\beta}$ line intensity ratio and the $[\rm O\,{III}]\lambda 5007$ flux density, F$_{[\rm O\,{III}]}$.
 These data are also presented in Table\,1.

 The X-ray data of our sample are collected from the literature, based on observations from {\it ASCA}, {\it BeppoSAX}, {\it Chandra}, and the
 {\it XMM-Newton} satellite. For some sources more than one value were reported in the literature for parameters such as the EW (K$_{\alpha}$) of the
 iron line and the observed X-ray flux, either due to intrinsic variability of the source or due to a different modeling of the spectra. For comparison,
 all results available from the literature are listed in Table\,1. For our statistical analyses, observational results were taken from
 {\it XMM-Newton} and {\it Chandra} whenever possible. Otherwise the most recent measurements were used.

 For the $[\rm O\,{III}]\lambda 5007$ emission line, the fluxes are mainly compiled from Dahari \& De Robertis (1988), Ho et al. (1997),
 Kim et al. (1995) and Whittle (1992). Again the most recent result was adopted when two or more observations were found. Since the NLR may be obscured
 by material in the host galaxy, the observed $[\rm O\,{III}]\lambda 5007$ flux should be corrected to determine its intrinsic flux. The narrow-line
 Balmer decrement was used to estimate this extinction. Assuming an intrinsic Balmer decrement $(H_{\alpha}/H_{\beta})_{0}=3$, the intrinsic
 $[\rm O\,{III}]$ line fluxes were derived from the formula $F_{[\rm O\,{III}],cor}=F_{[\rm O\,{III}],obs}[(H_{\alpha}/H_{\beta})_{obs}/(H_{\alpha}/
 H_{\beta})_{0}]^{2.94}$ (Bassani et al. 1999). The $[\rm O\,{III}]\lambda 5007$ fluxes were obtained for 46 maser sources and their luminosities were
 calculated from their extinction-corrected fluxes.

  The IRAS (Infrared Astronomical Satellite) point source catalogue was used to obtain the infrared fluxes (12, 25, 60, 100$\mu$m) for
  our H$_{2}$O maser host galaxies. Following the method of Wouterloot \& Walmsley (1986), the infrared flux ($6<\lambda<400$$\mu$m) was derived
  by extrapolating flux densities beyond 12 and 100$\mu$m and assuming a grain emissivity proportional to frequency $\nu$. In this way,
  infrared luminosities could be determined for 76 H$_{2}$O maser galaxies. Throughout the paper, the luminosity distance was derived using Calculators
  I provided by the NASA Extragalactic Database (NED), assuming $\Omega_{M}$ = 0.270, $\Omega_{vac}$ = 0.730, and $H_0$=70\,km\,s$^{-1}$\,Mpc$^{-1}$
  (e.g., Spergel et al. 2007).

\section{Results}

\begin{figure}
\begin{center}
 \includegraphics[width=10cm]{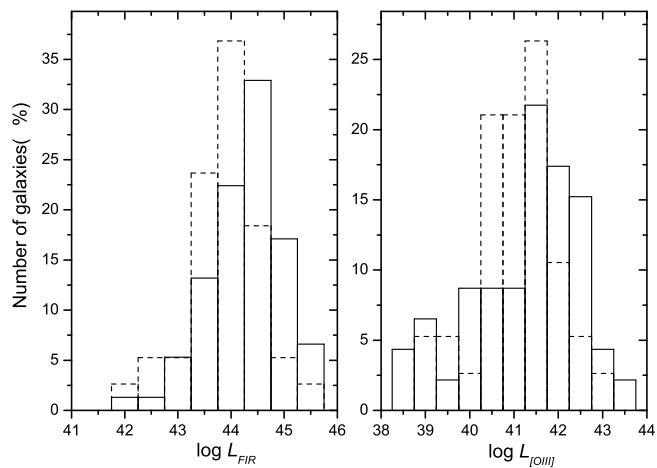}
\caption{Distributions of the FIR luminosity ($log\,L_{FIR}$, 76
sources, left panel) and the $[\rm O {III}]\lambda 5007$ line
luminosity ($log\,[O_{III}]$, 46 sources, right panel) for H$_{2}$O
maser sources (solid lines), in units of erg\,s$^{-1}$. For
comparison, the distributions are also given for a Seyfert 2 sample
without known H$_{2}$O maser emission  (dashed lines, 38 sources
from Mulchaey et al. 1994). All numbers are plotted on a percent
scale (\%).}
\end{center}
\end{figure}

\subsection{Isotropic indicators of the intrinsic nuclear power}

\begin{figure}
\begin{center}
 \includegraphics[width=12cm]{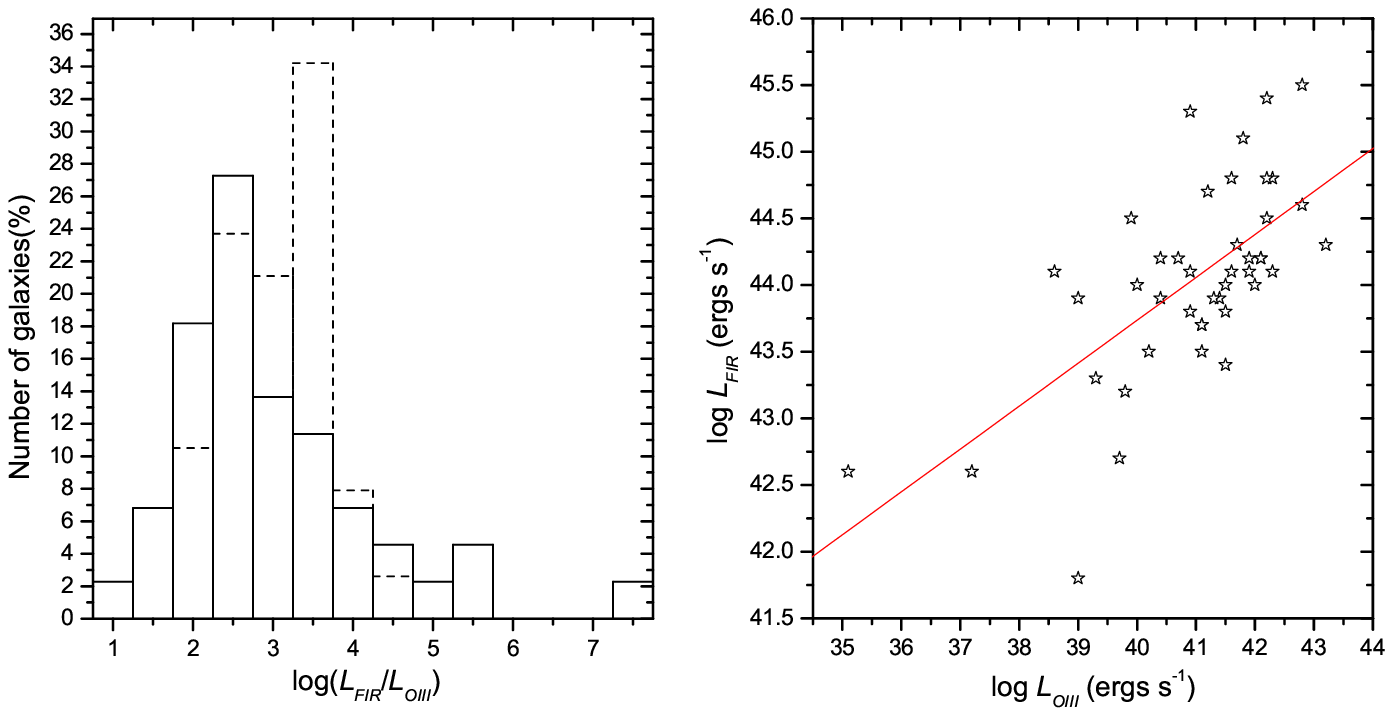}
\caption{Left panel: The distribution (on a percent scale) of the
ratio between the FIR and the $[\rm O\,{III}]$ line luminosity,
corrected for extinction, of H$_{2}$O maser galaxies on a
logarithmic scale. For comparison, the distribution is also given
for a Seyfert 2 sample without known H$_{2}$O maser emission (dashed
lines, 38 sources from Mulchaey et al. 1994). Right panel: the FIR
versus extinction-corrected $[\rm O\,{III}]$ luminosity. The
straight line shows a linear fit to the unweighted data.}
\end{center}
\end{figure}

  For our H$_{2}$O maser galaxies with available data, the luminosity distributions of both isotropic indicators, the $[\rm O\,{III}]$ line and the
  FIR luminosities, are presented in Fig. 1. The left panel shows the histogram of the infrared luminosities for all 76 H$_{2}$O maser
  galaxies with available data ($logL_{FIR}$, hereafter luminosity in logarithmic scale and in units of erg\,s$^{-1}$). $Log\,L_{FIR}$ ranges from 41.8
  to 45.5, with a mean value of 44.1$\pm$0.1 (the error denotes the standard deviation of the mean). For the sub-sample of 10 masers associated with
  massive star formation, the mean value of $<$$log\,L_{FIR}$$>$=43.3$\pm$0.2 is slightly fainter than that of the AGN masers. However, the difference
  between both distributions (not shown here) is not significant. The histogram shown in the right panel presents the number of the H$_{2}$O maser
  sources as a function of the $[\rm O\,{III}]$ line luminosity. It gives the range of $<$$log\,L_{[\rm O\,{III}]}$$>$ from 39 to 43.2 and a mean value
  of 40.9$\pm$0.3, comparable to results of Seyfert 2 galaxies not associated with detected H$_{2}$O maser (Mulchaey et al. 1994). For comparison, the
  latter sample is also presented in Fig. 1 (dashed lines).

  For the two tracers $Log\,L_{FIR}$ and $Log\,L_{[\rm O\,{III}]}$, the ratio was calculated and the distribution of the ratio is shown as
  a histogram in Fig. 2 (left panel). No significant difference can be found for the distribution of the ratio between our H$_{2}$O maser
  galaxies and the Seyfert 2 galaxy sample not containing galaxies with known masers (Mulchaey et al. 1994). This is in agreement with the fact that
  most H$_{2}$O maser sources have been found in Seyfert 2 systems. The good agreement between Seyfert 1 and Seyfert 2 galaxies (Mulchaey et al. 1994)
  with respect to this parameter, in spite of different viewing angles, lends further support to an approximately isotropic emission of both tracers.
  In order to further compare those two isotropic tracers, the FIR luminosity is plotted against the extinction-corrected
  $[\rm O\,{III}]$ luminosity of our maser sample in the right panel of Fig.\,2. A correlation is found, log\,L$_{FIR}$=(30.86$\pm$2.30)+
  (0.32$\pm$0.06)log\,L$_{[\rm O\,{III}]}$ with a Spearman's rank correlation coefficient r=0.66 and a chance probability $<$ 5$\times$10$^{-3}$.
  Assuming the corrected $[\rm O\,{III}]$ line luminosity to be a good isotropic tracer (Bassani et al. 1999), the strong correlation between
  FIR luminosity and $[\rm O\,{III}]$ line luminosity may suggest that the FIR luminosity is to some extent also an indicator of
  intrinsic nuclear activity, even though the FIR flux may be contaminated by a spatially extended starburst component.

\subsection{Optical thickness parameters and the 6.4\,keV iron line}


\begin{figure}
\begin{center}
  \includegraphics[width=16cm]{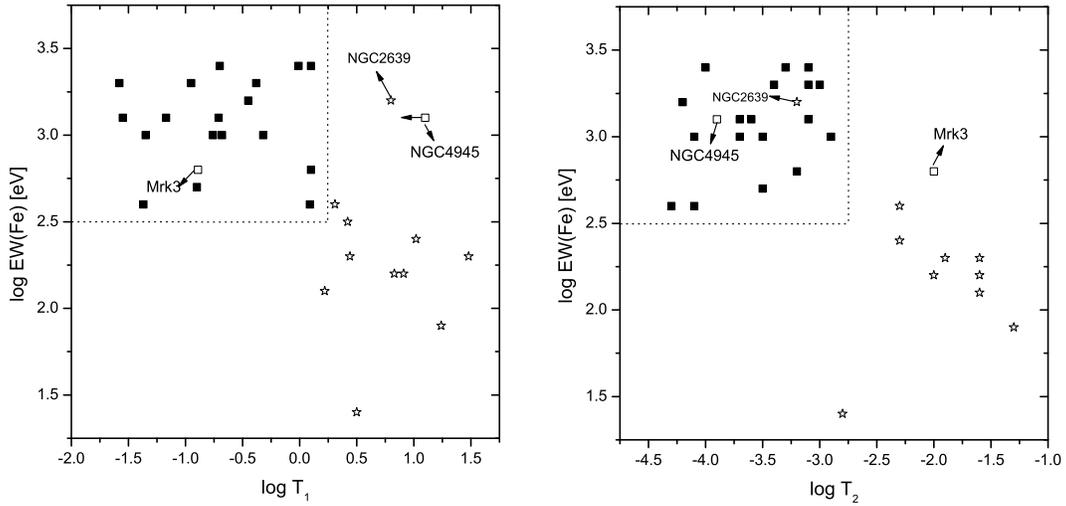}
 \caption{EW ($K_{\alpha}$) of the iron line versus the thickness parameters L$_{2-10keV}$/L$_{\rm [O\,{III}]}$ (T1, left panel) and
 L$_{2-10keV}$/L$_{FIR}$ (T2, right panel). Square and pentacle symbols represent Compton-thick and -thin sources as identified by modeling
 their X-ray spectra (e.g., Zhang et al. 2006). The bold squares show those Compton-thick sources, which have high EW (K$_{\alpha}$) and low T values.
 These are located in the upper-left regions of both panels. Three {\it exceptional} sources NGC\,2639, Mrk\,3, and NGC\,4945 are marked.
 For NGC\,4945, the lower limit of its [O\,III] luminosity is taken. Details are given in Sect.\,3.2.}
\end{center}
\end{figure}

 As mentioned above, nuclear absorbing columns can be obtained by analyzing X-ray spectra. These analyses are model dependent. Comparing the
 observed 2-10\,keV X-ray emission from the nuclear region, absorbed by the obscuring material along the line-of-sight, with the intrinsic nuclear
 power provides another method for evaluating the absorbing column density. The ratio of the observed 2-10\,keV luminosity and the luminosity of the
 nuclear isotropic indicators was assumed to represent the optical thickness parameter, which allows us to diagnose the gas absorption
 (Bassani et al. 1999). Here the L$_{X}$/L$_{\rm [O\,{III}]}$ and L$_{X}$/L$_{FIR}$ ratios are used as optical thickness parameters.
 In addition, high EW ($K_{\alpha}$) values of the iron emission line are considered as a qualitative feature indicating a heavily obscured
 nucleus (e.g., Maiolino \& Risaliti 2007). Combining the EW ($K_{\alpha}$) and the optical thickness parameters, we probe the circumnuclear
 environment of the maser sources.

 For the maser galaxies with available data (EW ($K_{\alpha}$), L$_{X}$, L$_{\rm [O\,{III}]}$ and L$_{FIR}$, in total 31 AGN maser sources),
 the EW (K$_{\alpha}$) of the iron lines is plotted against the optical thickness parameters in Fig.\,3. The prominent feature of the figure
 is the existence of an anti-correlation, which is similar to that obtained for the Seyfert 2 sample of Bassani et al. (1999), which is predominantly
 containing targets without known maser lines. In the left panel (EW ($K_{\alpha}$) v.s. L$_{X}$/L$_{\rm [O\,{III}]}$), a least-squares fit shows
 log\,(EW ($K_{\alpha}$))=(2.74$\pm$0.08)+ (-0.31$\pm$0.08)log\,L$_{X}$/L$_{\rm [O\,{III}]}$, with Spearman's rank correlation coefficient R=-0.57 and
 a chance probability of P $\sim$0.001. The possibly Compton-thick sources (shown by squares in Fig.\,3), determined by modeling their X-ray
 spectra, cluster in the upper left region, the region with high EW (K$_{\alpha}$) and low L$_{X}$/L$_{\rm [O\,{III}]}$ values. This area is marked
 by dashed lines. The approximate boundaries between Compton-thick and -thin sources, the latter shown by pentacles, are
 log\,(L$_{X}$/L$_{\rm [O\,{III}]}$)$\sim$0.25 (i.e., $L_{2-10\,keV}$$\sim$2\,$L_{\rm [O\,{III}]}$) and log\,EW ($K_{\alpha}$)$\sim$2.5 (i.e,
 EW$\sim$300\,eV). For comparison with Seyfert 2 galaxies without detected H$_{2}$O maser emission, the Seyfert 2 sample of Bassani et al. (1999) is
 used, excluding those objects with maser emission. The trend is the same for this sample (here not shown). Compton-thick sources fall again into the
 upper left region with high EW (K$_{\alpha}$) and low L$_{X}$/L$_{\rm [O\,{III}]}$ values, while Compton-thin sources are located in the lower
 right. The right panel of Fig.\,3 shows a similarly clear anti-correlation between EW (K$_{\alpha}$) and L$_{X}$/L$_{FIR}$. Linear fitting results
 in log\,(EW ($K_{\alpha}$))=(1.71 $\pm$ 0.26) + (-0.36$\pm$ 0.08)log\,L$_{X}$/L$_{FIR}$, with R=-0.61 and P $\sim$ 2 $\times$10$^{-4}$. Most of those
 sources with high EW (K$_{\alpha}$) and low L$_{X}$/L$_{FIR}$ value (upper-left region, in square symbols) are Compton-thick sources, while those
 sources with low EW (K$_{\alpha}$) and high L$_{X}$/L$_{FIR}$ value (in pentacles) are Compton-thin as determined from X-ray spectral fitting. The
 approximate boundaries between Compton-thick and -thin environments are in this case log\,L$_{X}$/L$_{FIR}$$\sim$-2.75 (i.e.,
 $L_{\rm FIR}$$\sim$600\,$L_{2-10\,keV}$) and log\,EW ($K_{\alpha}$)$\sim$2.5, i.e. again EW$\sim$300\,eV.

 We like to emphasize that these boundaries are not arbitrary. Among the 20 Compton-thick candidate sources from our sample, classified by
 conventional X-ray spectroscopy, we find 19 in each of our limited ``Compton-thick'' regions, related either to L$_{X}$/L$_{\rm [O\,{III}]}$ or to
 L$_{X}$/L$_{FIR}$. 18 of these sources are identical (see also Sect.\,4).

 The results obtained so far could be affected by systematic errors in the measurements. Assuming for the [O\,III] values uncertainties of
 $\sim$20\% (Dahari \& De Robertis 1988) results in L$_{X}$/L$_{\rm [O\,{III}]}$ errors of order 0.08\,dex. Obviously, this does not affect the robust
 fundamental trend in our diagnostic Fig.\,3. Compton-thick sources are still placed in the upper left and Compton-thin sources in the lower right. The
 anti-correlation between the EW ($K_{\alpha}$) and the optical thickness parameters are readily explained. With an increase of the absorbing column
 density, the X-ray luminosity will decrease so that L$_{X}$/L$_{\rm [O\,{III}]}$ and L$_{X}$/L$_{FIR}$ are reduced with respect to L$_{\rm [O\,{III}]}$
 and L$_{FIR}$. On the other hand, the EW ($K_{\alpha}$) values will increase, since these are measured against a reduced 6.4\,keV continuum level.

 With Compton-thick galaxies being located in the upper left part of the panels in Fig. 3, we find for our H$_{2}$O maser galaxies three criteria
 hinting at a Compton-thick nuclear environment: EW ($K_{\alpha}$)$>$300\,eV, $L_{2-10\,keV}$$<$2\,$L_{\rm [O\,{III}]}$ and $L_{\rm FIR}$$>$600\,
 $L_{2-10\,keV}$. These are independent of the detailed shape of the X-ray spectrum.

\subsection{Indicators for H$_{2}$O maser emission}

 The search for new extragalactic H$_{2}$O masers is ongoing and important with respect to several key aspects of modern astrophysics
 (see, e.g., Sect.\,1 and Braatz et al. 2009). Here we therefore investigate possible indicators of H$_{2}$O maser emission. We analyze relationships
 between H$_{2}$O maser luminosity and the iron line EW (K$_{\alpha}$), FIR, and ${\rm [O\,{III}]}$ luminosity.

\begin{figure}
 \begin{center}
  \includegraphics[width=8cm]{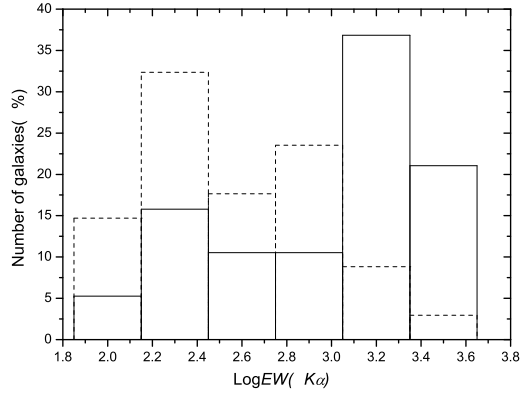}
 \caption{Distributions (on a percent scale) of the iron line EW ($K_{\alpha}$) (logarithmic scale, in units of eV) for Seyfert
 2s with detected H$_{2}$O maser emission (19 sources, solid lines) and non-masing Seyfert 2s (34 sources from Bassani et al. 1999, dashed lines).}
\end{center}
\end{figure}

 For 33 maser sources with available EW ($K_{\alpha}$) of the iron line in Table\,1, we obtain an EW (K$_{\alpha}$) mean value of
 $945 \pm 135$\,eV. X-ray observations show that strong iron line emission is common in the spectra of AGN hosting H$_{2}$O masers. It is
 interesting to check if the 6.4\,keV line can be used as a criterion to search for AGN masers. Have Seyfert 2s with detected maser emission higher
 EW (K$_{\alpha}$) values than non-maser Seyfert 2s? Our statistical results show that the iron line EW ($K_{\alpha}$) of masing Seyfert 2 galaxies
 (mean value $1063 \pm 169$\,eV and median value $\sim$800\,eV for our 19 maser sources) is higher than that of the non-maser Seyfert 2
 sample (mean value $375\pm60$\,eV and median value $\sim$200eV for 34 sources from Bassani et al. 1999 ). Figure 4 shows the distributions of both
 samples. While the difference seems to be obvious at first sight, we should cautiously avoid a definite conclusion due to the large scatter,
 the still too small number of sources, and the incompleteness of the studied samples. Potential differences in sensitivity have also to be addressed.
 For our maser host Seyfert 2s, the EW ($K_{\alpha}$) values were taken almost exclusively (except four sources) from \emph{XMM-Newton} or \emph{Chandra}
 observations. The results for Seyfert 2s without known masers (Bassani's sample) come mostly from ASCA data of lower sensitivity, which might lead to an
  increase in the real average (only sources with a strong iron line could be detected). This strengthens our result and amplifies the difference
  between Seyfert 2s with and without detected 22\,GHz H$_{2}$O maser. Nevertheless, we consider our result as tentative.

\begin{figure}
 \begin{center}
  \includegraphics[width=10cm]{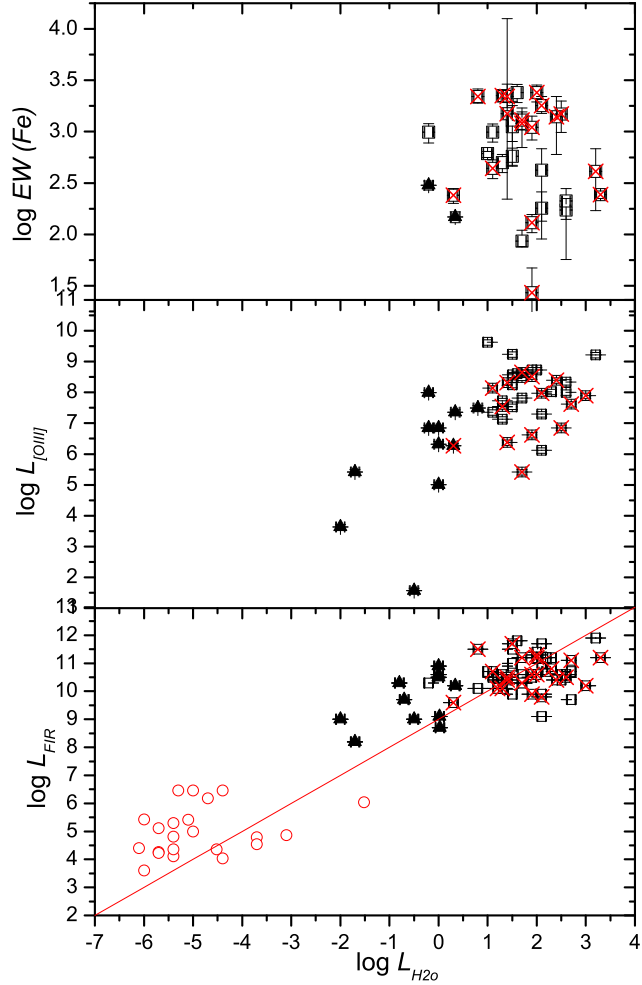}
 \caption{Upper panel: the EW ($K_{\alpha}$) of the iron line (logarithmic scale, in units of eV) versus isotropic luminosity of the H$_{2}$O
 maser emission (logarithmic scale, in $L_{\odot}$). Center: [O\,III] line luminosity against $L_{\rm H_2O}$; Bottom: FIR luminosity against
 $L_{\rm H_2O}$, empty circles show Galactic H$_{2}$O masers from Jaffe et al. (1981) and the line marks the correlation of
 $L_{\rm H_2O}$/$L_{\rm [FIR]}$$\sim$10$^{-9}$. Squares and triangles represent AGN-masers and masers in star formation regions
 respectively. Disk-masers, as a subsample of AGN-masers, for which maser emission is anticipated to be particularly well connected with indicators of
 the intrinsic nuclear power, are marked by crosses over squares.}
\end{center}
\end{figure}

 In order to investigate possible correlations between H$_{2}$O maser and iron emission lines, the EW ($K_{\alpha}$) of the Fe line was plotted against
 the isotropic H$_{2}$O luminosity in Fig.\,5 (upper panel). As already mentioned, H$_{2}$O maser emission can be produced by
 collisional pumping in a dense molecular layer, which is heated by irradiated X-rays from the nucleus (Neufeld et al. 1994).
 Strong Fe $K_{\alpha}$ emission is believed to be produced via X-ray reflection by the cold iron in the circumnuclear region
  (e.g., Fabian et al. 2000). In those cases where H$_{2}$O maser and iron line emission are detected, the nuclear X-ray emission plays a key role.
  Correlations between maser and iron line emission are therefore expected. However, our results show no apparent trend between the
  iron line EW (K$_{\alpha}$) and the isotropic H$_{2}$O maser luminosity, neither for the entire sample nor the two subsamples, AGN- and
  star-forming masers (squares and triangles, respectively, in Fig.\,5). In view of alternative H$_{2}$O excitation mechanisms (see, e.g., Lo
  2005), the subsample of possible disk-masers (with detected high velocity maser features) was analyzed separately (see the crosses in Fig.\,5).
  However, even in this case no significant correlation can be found.

 The [O\,III] and FIR luminosities were also plotted against H$_{2}$O maser luminosity in Fig.\,5 (central and bottom panel
 respectively). There is no significant correlation between $L_{\rm [O\,III]}$ and $L_{\rm H_2O}$, although maser sources related to
 star formation seem to have lower [O\,III] luminosities than AGN masers. For FIR versus H$_{2}$O maser luminosities, there appears to be
 a correlation, similar to that previously found by Henkel et al. (2005), Castangia et al. (2008), Bennert et al. (2009) and Surcis et al.
 (2009). The relation was first found for galactic star forming regions by Jaffe et al. 1981, i.e., luminous H$_{2}$O masers form in star
 formation regions with high FIR luminosity. For comparison, values of Galactic H$_{2}$O masers are also plotted in Fig.\,5 (empty circles)
 and the line shows the correlation of $L_{\rm H_2O}$/$L_{\rm [FIR]}$$\sim$10$^{-9}$ from Jaffe et al. (1981). Apparently, there exists a correlation
 between FIR and H$_{2}$O maser luminosity over many orders of magnitude. When considering AGN-masers only (squares in Fig.\,5, lowest panel), the
 strongest masers appear to be overluminous with respect to the $L_{FIR}$-$L_{H_{2}O}$ correlation. This is likely caused by the different properties
 of AGN versus star-forming masers.

\section{Discussion}

 Compton-thick nuclei are known to contribute a significant fraction of the hard X-ray background. Their density as a function of redshift is also a
 relevant parameter when studying the evolution of the universe. Thus our newly defined criteria identifying such nuclei may be helpful when trying to
 reach this goal.

 Based on our new approach introduced in Sect.\,3.2, 18 H$_{2}$O maser sources in the upper left part of the panels in Fig.\,3
 (see the solid squares) can be considered to be Compton-thick. These are NGC\,591, NGC\,1068, Mrk\,1066, NGC\,1386, NGC\,2273, UGC\,5101,
  NGC\,2782, NGC\,3079, IC\,2560, NGC\,3393, Arp\,299, NGC\,5194, Mrk\,266, NGC\,5347, Circinus, NGC\,5643, NGC\,5728 and NGC\,6240. Comparing this with
 Zhang et al. (2006), there are five new sources, NGC\,591, UGC\,5101, NGC\,3072, IC\,2560, Mrk\,266. About 60\% of the 31 AGN masers turn out to be
 Compton-thick. This is consistent with the result found by Greenhill et al. (2008). For maser sources associated with Seyfert 2 nuclei, 60\% (15/25) are
 Compton-thick, which is, however, not significantly, higher than that of Seyfert 2 objects without detected maser emission ($\sim$45\%, 9/20 from
 Risaliti et al. 1999; 19/42 from Bassani et al. 1999).

 Our sample also contains a few \emph{exceptional} sources, which show the limits of our selection criteria. For NGC\,4945, its
 low [O\,III] line flux places the source outside our Compton-thick border (EW ($K_{\alpha}$)-$L_{X}$/$L_{\rm [O\,{III}]}$, see left panel in Fig.\,3).
 However, the [O\,III] flux from Risaliti et al. (1999) only gives a very stringent lower limit to the actual intrinsic [O\,III] emission. The
 lack of a reliable [O\,III] flux is thought to be due to high absorption in the edge-on galactic disk, instead of its intrinsic weakness. This is
 supported by its hard X-ray spectrum, which indicates that NGC\,4945 hosts one of the brightest AGN in the hard X-ray range ($>$100\,keV) and can
 therefore be considered to contain a `bona-fide' Compton-thick nucleus (e.g., Guainazzi et al. 2000b). Similar to NGC\,4945, Mrk\,3 is also considered
 to be a `Bona-fide' Compton-thick Seyfert 2 from its large brightness in the hard X-ray range (e.g., Cappi et al. 1999). However, the source is
 found outside our limited EW ($K_{\alpha}$)-L$_{X}$/L$_{FIR}$ region (right panel in Fig.\,3) for its relatively low value of the infrared flux. More
 constraints are desirable to probe its circumnuclear environment. The disk-maser galaxy NGC\,2639 is located inside the required EW
 ($K_{\alpha}$)-L$_{X}$/L$_{FIR}$ region and outside the EW ($K_{\alpha}$)-L$_{X}$/L$_{\rm [O\,{III}]}$ area. However, large uncertainties in
 the X-ray results (ASCA observations) have to be noted. NGC\,2639 is a weak X-ray source with an ASCA count rate $<$0.01\,counts\,s$^{-1}$. The ASCA
 data were analyzed by Wilson et al. (1998) and Terashima et al. (2002) and the source was considered to be Compton-thin ($N_{\rm H}$$\sim$5$\times$10
 $^{23}$\,cm$^{-2}$). Low photon statistics lead to uncertainties in the fitting models and Chandra observation are therefore needed to investigate its
 highly obscured nucleus.

 We conclude with some cautionary notes. First, uncertainties are involved when using the $[\rm O\,{III}]\lambda 5007$ line and FIR luminosity
 as nuclear isotropic indicators. Although assumed to be isotropic, the $[\rm O\,{III}]$ line luminosity might depend on the geometry of the
 system, for example on the opening angle of the torus and the inclination of the large scale disk. Shielding effects may affect the ionizing
 radiation seen by the NLR. FIR emission of H$_{2}$O megamaser galaxies may mainly arise from the AGN environment, but there is possible
 contamination from a starburst component and it is unclear how much it contributes. Second, the EW (K$_{\alpha}$) of the iron line can be affected by
 other factors, such as the geometry of the accretion disk and the inclination angle at which the reflecting surface is viewed (e.g., Fabian et al. 2000
 , Bianchi et al. 2005). A high EW (K$_{\alpha}$) of the iron line can also appear if the radiation is anisotropic or if there is a time lag between a
 drop in the continuum and the line emission (Bassani et al. 1999). With future advanced X-ray telescopes, sensitive observations of more H$_{2}$O maser
 host galaxies at higher energies (above 10\,keV) will provide important complementary information, further constraining nuclear column density.

\section{Summary}

 In this paper, multi-wavelength data from the complete sample of galaxies ($D$$>$100\,kpc) so far reported to host 22\,GHz H$_{2}$O masers are
 analyzed, including the equivalent width (EW) of the iron $K_{\alpha}$ line, the $[\rm O\,{III}]\lambda 5007$ line, and the X-ray and FIR
 emission. The latter two are considered to be the isotropic tracers of intrinsic nuclear power. The observed nuclear X-ray luminosity, compared with the
 luminosities of these two isotropic tracers (our optical thickness parameters), can be used as a measure of the circumnuclear absorption. The EW
  ($K_{\alpha}$) and the optical thickness parameters ($L_{X}$/$L_{[\rm O\,III]}$, $L_{X}$/$L_{FIR}$) are combined here to probe the obscuration of
  maser host AGN. The main results are summarized below:

 (1) Our statistical analysis shows obvious anti-correlations between the EW ($K_{\alpha}$) of the Fe emission line and the two optical
 thickness parameters. Without requiring a full X-ray spectrum, Compton-thick nuclear environments can be identified with these parameters and are found
 to be characterized approximately by EW ($K_{\alpha}$)$>$300\,eV, $L_{2-10\,keV}$$<$2\,$L_{[\rm O\,{III}]}$ and $L_{\rm FIR}$$>$600\,$L_{2-10\,keV}$;

 (2) 18 H$_{2}$O maser sources matching these criteria are identified to be Compton-thick. A comparison with Zhang et al. (2006) shows,
 that among these there are five newly identified H$_{2}$O maser galaxies which are Compton-thick, i.e., NGC\,591, UGC\,5101, NGC\,3072, IC\,2560 and
 Mrk\,266. Masers associated with Seyfert 2 nuclei may be more likely Compton-thick (60\%) than Seyfert 2s without detected maser emission
 ($\sim$45\%). However, this still has to be confirmed with larger samples;

 (3) In an attempt to guide future H$_{2}$O maser surveys, new ways to find extragalactic H$_{2}$O sources are also explored. H$_{2}$O maser
 sources may show larger EW (K$_{\alpha}$) values than non-maser Seyfert 2s, which, however, also needs further support. No significant correlations
 have been found between EW (K$_{\alpha}$), $L_{[\rm O\,III]}$ and $L_{\rm H_2O}$. There appears a linear correlation between $L_{FIR}$ and
 $L_{\rm H_2O}$, which is consistent with the correlation found for Galactic H$_{2}$O masers. However, the strongest H$_{2}$O masers appear
 overluminous with respect to their $L_{FIR}$. This may be related to their different origin when compared with masers associated with sites of massive
 star formation well outside the nuclear region of their parent galaxy.

\acknowledgments

 We wish to thank the anonymous referee for many detailed and constructive comments as well as P. Castangia for critically reading the
 manuscript. This work is supported partly by the National Natural Scientific Foundation of China (10633010) and GuangDong province Natural Science
 Foundation (8451009101001047). We made use of the NASA/IPAC extragalactic Database (NED), High-Energy Astrophysics Science and Research Center (HEASARC)
 and the NASA Astrophysics Data System Bibliographic Services (ADS).

\label{lastpage}

\end{document}